\documentclass[runningheads]{llncs}

\usepackage{algorithm}
\usepackage{algpseudocode}
\usepackage{graphicx}
\usepackage{amssymb}

\usepackage{wrapfig}

\begin{document}

\title{Runtime Verification of Linux Kernel Security Module\thanks{This work has received funding from the Ministry of Education and Science of Russia under grant agreement RFMEFI60719X0295.}}
\author{Denis Efremov \and Ilya Shchepetkov}

\institute{
ISP RAS, Moscow, Russia\\
\email{\{efremov,shchepetkov\}@ispras.ru}
}

\maketitle

\begin{abstract}
The Linux kernel is one of the most important Free/Libre Open Source Software (FLOSS) projects.
It is installed on billions of devices all over the world, which process various sensitive, confidential or simply private data.
It is crucial to establish and prove its security properties.
This work-in-progress paper presents a method to verify the Linux kernel for conformance with an abstract security policy model written in the Event-B specification language.
The method is based on system call tracing and aims at checking that the results of system call execution do not lead to accesses that violate security policy requirements.
As a basis for it, we use an additional Event-B specification of the Linux system call interface that is formally proved to satisfy all the requirements of the security policy model.
In order to perform the conformance checks we use it to reproduce intercepted system calls and verify accesses.

\end{abstract}
\section{Introduction}
Access control mechanisms in operating systems are usually implemented based on a security policy model, which contains description of the security properties to be enforced by these mechanisms.
A security policy model may be a simple text document, but for a certain level of assurance it should be formalized and verified, as stated by the Common Criteria standard~\cite{cc1,cc2}.
An additional level of assurance may be achieved by demonstrating that the implementation of an access control mechanism indeed conforms to its formal specification.

Access control mechanisms in Linux are implemented in the kernel.
We propose to intercept system calls to the kernel while performing various actions like creating, reading, writing, deleting files, spawning processes, etc, and check that the results of their execution do not lead to accesses that are forbidden by the security policy model.
It is difficult to check directly because of the abstraction gap.
Security policy models are often too high-level comparing to concrete data structures and functions of the Linux kernel.
To overcome the difference between the specification and the implementation we develop an Event-B~\cite{abrial_modeling_2010} specification of the Linux system call interface, formally prove that it satisfies all requirements of the security policy model (which is also formalized in Event-B), and then translate it to an executable form which is more suitable for checking correctness of intercepted system calls.

The following section briefly describes the security policy model in use.
Section 3 depicts the Event-B language in which the model was formalized and verified.
Section 4 briefly describes the formal specification of the security model.
Section 5 provides a description of an additional Event-B specification required to perform a conformance verification.
Section 6 describes the Linux security modules framework, which is used to implement security policy models inside the kernel.
Section 7 presents the runtime verification method itself.
Related work is observed in Section 8.
The final section concludes the paper and considers future work.

\section{Security Policy Model}
A security policy is a high-level specification of the security properties that a given system should possess, and of security mechanisms that enforce those properties.
Security policies are described in the form of security policy models as \textit{state transition systems}, where each possible state transition from a \textit{secure} state must preserve security properties and produce another secure state.
The state is declared secure if all current accesses and permissions are in accordance with a security policy.

Operating system (OS) security policy models define the rules for controlling accesses of subjects (users and programs running on their behalf) to various objects (files, directories, devices) and other subjects.
They define state transitions as transition functions that model usual OS actions, like creating and deleting files, processes, requesting accesses, etc.
Examples of such models would be the classic Bell-LaPadula~\cite{bell_secure_1973,bell_secure_1976} and Biba~\cite{biba_integrity_1977} models, which were first to describe semantics and security properties of multilevel security and mandatory integrity control respectively.

In this paper we use a Hierarchical Integrated Model of Access Control and information Flows (the HIMACF model, previously known as the MROSL DP-model~\cite{devyanin_book_2013,devyanin_formal_2014}).
It describes means to enforce the separation of information based on confidentiality and integrity requirements.
It combines several security mechanisms:
\begin{itemize}
\item Role Based Access Control (RBAC).
In RBAC, permissions to perform various actions are grouped intro roles and are assigned to a user by an administrator or obtained through special administrative roles. RBAC is often used as a replacement for more simple discretionary access control;
\item Mandatory Integrity Control (MIC).
In MIC, an integrity level is assigned to all users, processes and files.
That level represents their level of trustworthiness, so the higher the level~--- the more trusted and important a user, a process or a file.
MIC controls accesses of subjects to objects according to their integrity levels.
MIC is implemented in Windows and macOS to protect system files from modification by users or malicious software;
\item Multilevel Security based on Mandatory Access Control (MLS, MAC).
It was designed to deal with classified documents in military computer systems.
MLS controls accesses according to the user's clearance and the file's classification.
\end{itemize}

\begin{wrapfigure}{r}{0.35\textwidth}
    \vspace{-30pt}
    \begin{center}
        \includegraphics[width=0.33\textwidth]{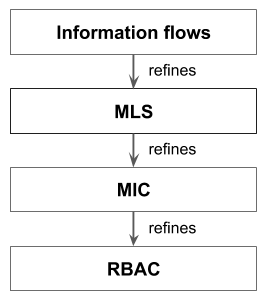}
    \end{center}
    \vspace{-15pt}
    \caption{The hierarchy of levels in the HIMACF model.\label{fig:himacf}}
\end{wrapfigure}

These mechanisms are integrated into a linear hierarchy, where each next level is based on the previous ones.
Also since the sequence of perfectly normal and secure accesses may lead to insecure information flows, there is an additional level that contains proofs of their absence (see Fig.~\ref{fig:himacf}).

The HIMACF model is implemented in the certified distribution Astra Linux Special Edition~\cite{astra_se} using the Linux Security Modules framework~\cite{lsm}.
The model is written in plain text with extensive use of math and consists of approximately 300 pages.
We have formalized and verified~\cite{shchepetkov_refinement_2016} it using the Event-B specification language.
It took us 4 years, and during this process we have found and fixed a number of issues and inconsistencies in the HIMACF model.

\section{Event-B}
Event-B is a formal method based on set theory and predicate logic.
It has a simple notation and comes with a tool support in the form of the Rodin Platform~\cite{abrial_rodin_2010}.
It is mainly used for developing models of various control systems, but it is also particularly well suited for security policy modeling.

An Event-B specification is a \textit{discrete transition system} and consists of \textit{contexts} and \textit{machines}.
Contexts contain the static, or unchanged parts of the specification: definitions of \textit{carrier sets}, \textit{constants}, \textit{axioms}.
Machines contain the dynamic or behavioral parts of the specification: \textit{variables}, \textit{invariants} and \textit{events}.

Event-B is a state-based method, so values of variables form the current state of the specification.
Events represent the way the state changes over time --- the transition.
Events may contain \textit{parameters}, \textit{guard conditions} that are necessary for the event to be enabled (or preconditions), and \textit{actions}, that change variables' values.
Invariants describe important properties of the system and are supposed to hold whenever variable values change, so such changes need to be explicitly proven to be correct.
For each case that requires a proof the Rodin platform generates a corresponding \textit{proof obligation}, that can be discharged automatically using various provers and solvers or interactively.
Interactive proofs are also automatically checked for soundness.

\section{Event-B Specification of the HIMACF Model}
The HIMACF model uses set theory and predicate logic for defining the state and the properties that the state must satisfy, and it also contains several atomic state transition rules which describe events taking place in the operating system.
It makes its structure very similar to the structure of a typical Event-B specification, so its formalization in Event-B was quite straightforward\footnote{Publicly available part of the specification: https://github.com/17451k/base-model}.

The state variables of the Event-B specification are expressed as sets and functions (set of ordered pairs with additional restrictions):
\begin{itemize}
\item Sets:
\begin{itemize}
\item user accounts;
\item entities (objects and containers);
\item subjects;
\item roles (administrative, ordinary, negative).
\end{itemize}
\item Functions:
\begin{itemize}
\item integrity and security levels (in the form of lattice);
\item current accesses and access rights (or permissions) to entities and roles;
\item hierarchies of roles, entities and subjects;
\item some additional relations between elements of the specification;
\item various flags.
\end{itemize}
\end{itemize}

These variables describe the usual operating system elements like user accounts, subjects (which are processes), entities (files, directories, sockets, etc.), and roles.
Each of these elements have integrity and security labels that are mapped to them by a number of corresponding functions.
Some additional things are also modelled as functions, like current accesses, permissions, hierarchies, and so on.

In total the specification contains 65 state variables.
There are also 80 events that describe possible state transitions typical for an OS:
\begin{itemize}
\item Create or delete entities, user accounts, subjects, roles;
\begin{itemize}
\item create or delete hard links for entities and roles;
\item rename entities or roles;
\end{itemize}
\item Get or delete accesses, access rights to roles, entities;
\item Change security, integrity labels, various flags;
\item Additional events for analysis of information flows;
\begin{itemize}
\item example: if an entity \textit{x} have write access to a subject \textit{y}, which have write access to a subject \textit{z}, then there can be an information flow from \textit{x} to \textit{z}.
\end{itemize}
\end{itemize}

Finally, the specification contains 260 invariants divided into three groups.
First one are type invariants: they describe types of all state variables.
For example, the type of the variable that contains accesses of subjects to entities is expressed in Event-B like this:
$SubjectAccesses \in Subjects \rightarrow (Entities \leftrightarrow Accesses)$.
Another group is consistency invariants: they impose correctness constraints on the system state.
For instance, if we have a variable that describes filesystem (hierarchy of files and folders), then it must not contain cycles, i.e., a folder cannot contain itself, even indirectly.

The last group of invariants is the most important one: it contains all security properties of corresponding security mechanisms.
For example, there is the following security property: if a subject has write access to an entity, then its integrity label must be greater or equal than the integrity label of this entity.
It is expressed in Event-B like this: $\forall s, e \cdot s \in Subjects \land e \mapsto WriteA \in SubjectAccesses(s) \Longrightarrow EntityInt(e) \leqslant SubjectInt(s)$.

\section{Event-B Specification of the System Call Interface}

The HIMACF model and its Event-B specification, however, are quite abstract and different from the concrete data structures and functions of the Linux kernel, which contain the security policy implementation as the Linux Security Module.
To prove their conformance it is necessary to reduce this gap.
Event-B supports \textit{the refinement technique}~\cite{abrial_refinement_2007} to represent systems at different abstraction levels that can be used to resolve this issue.

We have used refinement to develop an additional Event-B specification of the Linux kernel system call interface.
Using Rodin we have formally proved that the additional specification correctly refines the Event-B specification of the HIMACF model and thus satisfies its properties.
Hence, if we will show the conformance between the additional specification of the system call interface and the Linux kernel, then the desired conformance between the Linux kernel and the security policy model will be derived automatically.

The additional specification, however, has quite an unusual structure.
The difference lies in the nature of system calls: the exact sequence of actions that will be performed as the result of the system call depends on the current state of the OS and on the arguments of the call.
Because of this variability it is impossible to model them as single atomic events.
Instead, we used a different approach.

To overcome this issue we have decided to represent each system call as a \textit{graph of events} connected together with the special state variable called \textit{Next}.
\textit{Next} is used to specify the order in which normally independent events should occur.
This is achieved as follows: each event in the graph of events have a guard condition specifying that it can only occur if the current value of the \textit{Next} variable is the name of this event.
Depending on other guards the event also changes the value of the \textit{Next} variable with the name of the event that should follow next.

Each graph of events representing a system call have a single entry node (the ``initial'' event), a single exit node (the ``final'' event) and a large amount of paths in between.
Each path is a series of events and the next event in the path is specified by the current value of the \textit{Next} state variable.
The path (concrete series of events representing a particular execution of the system call) is defined by the parameters of the ``initial'' event and the current state of the specification in a way that for each event in the path there is no more than one possible next event.

Let's consider the \texttt{open()} system call to open or create a file.
This system call has the following declaration\footnote{According to the Linux manual page\\ http://man7.org/linux/man-pages/man2/open.2.html}: \texttt{int open(const char *pathname, int flags)}.
The \texttt{open()} call has two arguments: \texttt{pathname} specifies the file to open, and \texttt{flags} determines its access mode: read-only, write-only, or read/write.
These access modes are expressed by corresponding flags \texttt{O\_RDONLY}, \texttt{O\_WRONLY} and \texttt{O\_RDWR}.
\texttt{flags} may also contain additional file creation and status flags.
The return value of \texttt{open()} is a file descriptor, which can be later used in subsequent system calls (\texttt{read()}, \texttt{write()}, etc.).

Now let's consider a specific case of \texttt{open()} system call in which the file from the \texttt{pathname} argument does not exist, and the \texttt{flags} argument contains \texttt{O\_WRONLY} (open file to write) and \texttt{O\_CREAT} (create file if it does not exist) flags.
If the process which calls \texttt{open()} has all necessary permissions, then \texttt{open()} performs the following sequence of actions:

\begin{itemize}
    \item parse and validate values of it arguments;
    \item check that the process has all necessary permissions. In this case the check is successful;
    \item get the process write access to the directory where the file will be created;
    \item create the file;
    \item get the process permission to write to the created file;
    \item get the process write access to the created file;
    \item return file descriptor of created and opened to write file.
\end{itemize}

This case can be formalized in the Event-B specification of the system call interface as the sequence of 8 events: \texttt{open\_start, open\_check\_p, open\_write\_p, open\_create, open\_grant, open\_check, open\_write, open\_finish}, where:
\begin{itemize}
    \item \texttt{open\_start} contains preconditions (guards) that analyze arguments of the \texttt(open) call and decide which event should occur next. In the given case, the file being opened does not exist, so the next event is \texttt{open\_check\_p}. If the file existed, the next event would be \texttt{open\_check}, and the sequence of events would be different;
    \item \texttt{open\_check\_p} checks that the process has all necessary permissions. In this case the check is successful, so the next event is \texttt{open\_write\_p};
    \item \texttt{open\_write\_p} is a refinement of the \texttt{access\_write\_entity} event of the Event-B specification of the HIMACF model. This event grants the process write access to the directory where the file will be created;
    \item \texttt{open\_create} is a refinement of the \texttt{create\_object} event of the Event-B specification of the HIMACF model. This event creates the file;
    \item \texttt{open\_grant} is a refinement of the \texttt{grant\_rights} event of the Event-B specification of the HIMACF model. This event grants the process permission to write to the created file;
    \item \texttt{open\_check} checks that the process has necessary permissions to obtain access to the created file and decides which event should occur next. In this case the process opens file to write, so the next event is \texttt{open\_write};
    \item \texttt{open\_write} is a refinement of the \texttt{access\_write\_entity} event of the Event-B specification of the HIMACF model. This event grants the process write access to created file;
    \item \texttt{open\_finish} returns the requested file descriptor.
\end{itemize}

\begin{figure}
    \begin{center}
        \includegraphics[width=0.6\textwidth]{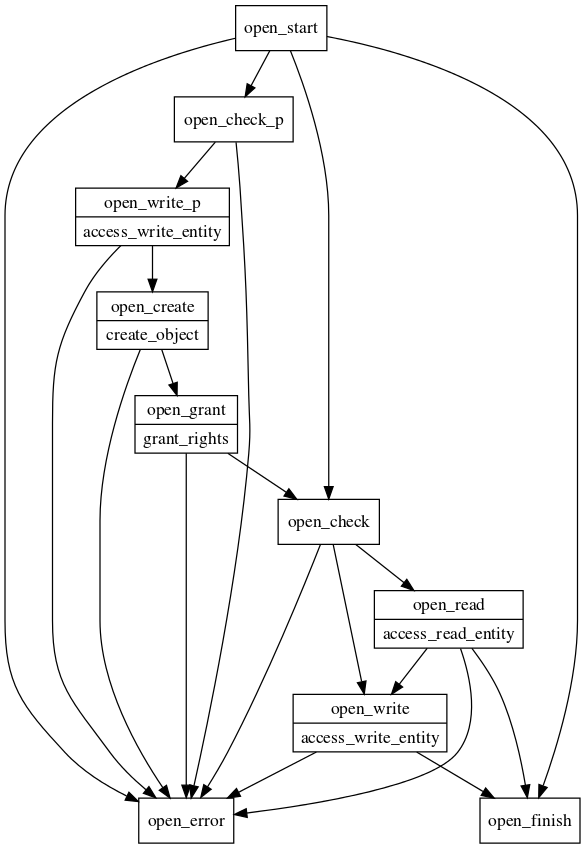}
    \end{center}
    \caption{Graph of events corresponding to several special cases of open() system call.\label{fig:open}}
\end{figure}

This sequence of events corresponds to one specific case of \texttt{open()} system call.
To demonstrate our approach we have formalized a few more cases\footnote{Code can be found here: https://github.com/17451k/base-model/tree/open}(see Fig.~\ref{fig:open}).
You can see that the graph consists mostly from the same events, but there are more possible paths between them.

If we formalize all the remaining cases, the resulting graph will be a formal specification of the behavior of the \texttt{open()} system call.
Due to the use of refinement, this specification will be correct by construction and fully conform to the rules and events of the HIMACF model.
In turn, this will mean that for any combination of parameters and the state of the system, executing the \texttt{open()} system call will hold all the security properties of the HIMACF model.

\begin{figure}
    \begin{center}
        \includegraphics[width=0.7\textwidth]{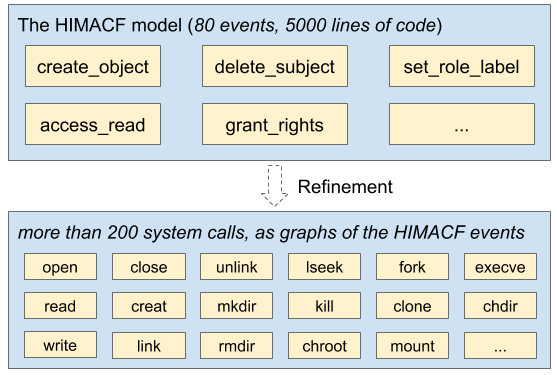}
    \end{center}
    \caption{Refinement between Event-B specifications of the HIMACF model and the system call interface.\label{fig:essci}}
\end{figure}

All system calls can be formalized in a similar way, resulting in the specification of the system call interface that is proved to be consistent and complete.
But such specification can turn out to be extremely large (several times more than the Event-B specification of the HIMACF model) and difficult to write and prove, mainly from the complicated refinement relation between them (see Fig.~\ref{fig:essci}).

\section{Linux Security Modules}
\begin{figure}
    \begin{center}
        \includegraphics[width=1\textwidth]{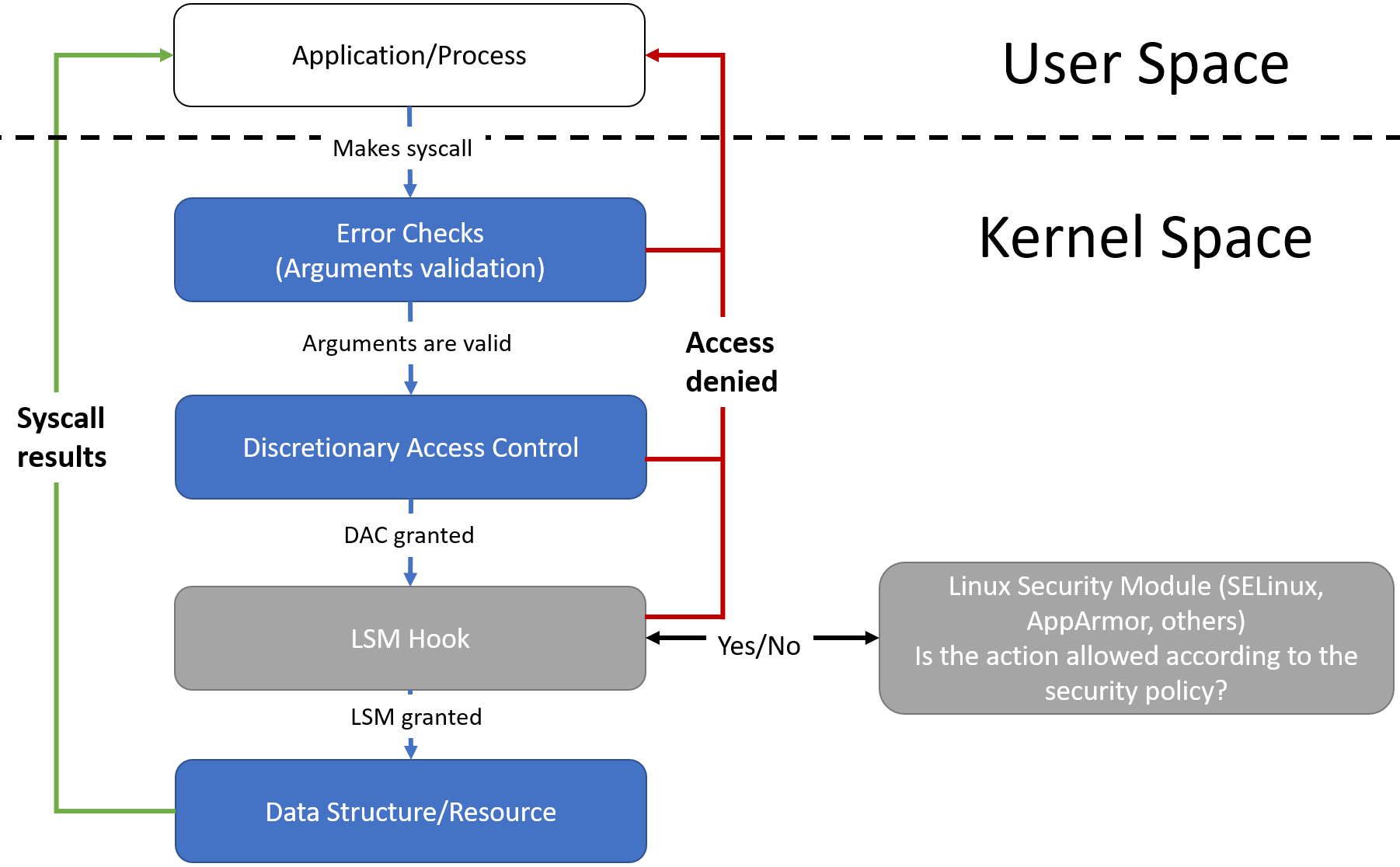}
    \end{center}
    \caption{Linux Security Modules (LSM) hooks.\label{fig:lsm_diagram}}
\end{figure}

In Linux, userspace programs work with external resources via the kernel, and make requests for accesses through system calls.
When a program executes a system call to, for example, open a file, the kernel performs a number of checks. It verifies the correctness of the passed arguments, checks the possibility of their allocation, and also checks that the program has the permission to obtain the requested resource by the discretionary access control.
If more advanced access control mechanisms are used, then the kernel also checks that the access request satisfies their security policy.
Such mechanisms are called security modules and based on the Linux Security Modules (LSM) framework.
LSM adds a call to a security module after the discretionary access checks in a control flow of a system call handling.
These calls are placed across the kernel and called LSM hooks~(see~Fig.~\ref{fig:lsm_diagram}). 

There are several potential cases when the kernel manages permissions and accesses incorrectly.
First, it is possible that the control flow does not reach a security module~\cite{write_lsm_mh_2005,georget_lsm_mh_2016,viro_lsm_mh_2006}.
The LSM interface may not be complete enough, so it may lack hooks to check certain situations~\cite{jurgens_lsm_infiniband,goyal_lsm_overlayfs}.
A security module can also be implemented incorrectly and grant accesses that should not be granted, or deny accesses that should be granted.
There is always place for errors due to the abstraction gap and specifics of kernel~--- module interactions.
Thus, we want to verify that the kernel of Astra Linux distribution with the security module indeed conforms the Event-B specification of the HIMACF model and this includes all enumerated errors.

It is worth to note that the Event-B specification of the system call interface does not model such things as the availability of resources (number of processes, virtual memory), and does not contain description of the discretionary access control mechanism.
So, for example, the kernel could deny an access due to the lack of physical resources of the machine, but at the same time the specification grants it assuming that the resources are unlimited.
Thus the divergence between the specified behavior and the real one should be treated as an error only in case the security policy model denies the access, but the security module grants it.

\section{Runtime Verification Method}
We propose to demonstrate the absence of such divergences by means of runtime verification, which require a test suite.
The test suite should cover various patterns of access requests.
In this paper we do not consider the issue of constructing tests and instead use special tests for our model and whole system tests such as Spruce~\cite{spruce}, ltp~\cite{ltp} fuzzing with syzkaller~\cite{syzkaller}.
These test suites allow us to achieve relatively good line coverage (more than 80\%) on our security module and to cover all LSM hooks in target subsystems of the Linux kernel. 

The runtime part of the method is divided into two consecutive steps: gathering of information about the kernel behavior (monitoring) and its analysis.

At the first step, the execution traces of the Linux kernel are collected.
It is performed while a test suite is run.
In order to reproduce such traces on the specification we also need to record a global state of the kernel, which is performed at the very beginning of this step.
This includes, for example, information about running processes, opened files, shared resources, etc.

Traces contain arguments of the system call and the result of its processing by the kernel (output arguments and the result code).
Along with this, each trace contains an additional information that is necessary for mapping the global kernel state to the state of the specification, such as inodes and dentries for files, user ids, etc.

We use SystemTap~\cite{jacob2008systemtap} tool to gather the traces from the kernel. It allows one to describe desired probe points in the kernel, such as system calls, with a special language and log the state of in-kernel data structures to a journal. 

\begin{algorithm}[htb]
    \caption{Replay of single system call on the Event-B specification}\label{lst:syscall_replay}
    \begin{algorithmic}[1]
      \Procedure{Replay\_Syscall}{spec, syscall}
      \State $syscall\_graph := spec[syscall[name]]$
      \State $params := syscall[args]$
      \State $event := syscall\_graph[initial]$ 
      \While{$event \neq syscall\_graph[final]$}
          \If{$guards\_hold(spec[state], event, params)$}
          \State $spec[state] \gets event(spec[state], params)$\Comment{update}
          \Else
          \State \textbf{return} $Denied$
          \EndIf
          \State $event := next(spec[state], event, params)$
      \EndWhile
      \State \textbf{return} $Granted$
      \EndProcedure
  \end{algorithmic}
\end{algorithm}

At the second step, we initialize the state of the Event-B specification with the state of the kernel and replay system calls from the trace on the Event-B specification.

The replay algorithm consists of the following steps~(see~Alg.~\ref{lst:syscall_replay}):
\begin{enumerate}
    \item Pass the arguments of the system call as parameters to the ``initial'' event of its specification;
    \item Check that all guards of the current event are satisfied. If they are not satisfied, then report that the access is denied according to the security policy rules~(lines~6,~9);
    \item If current event is not ``final'', then compute the ``next'' event, apply event to the current state to change it~(line~7), mark the ``next'' event as current and return to step 2;
    \item If current event is ``final'', then apply it to the current state and report that the access is granted according to the security policy rules;
\end{enumerate}

\begin{algorithm}[htb]
    \caption{Replay of kernel traces on the spec}\label{lst:trace_replay}
    \begin{algorithmic}[1]
    \Procedure{Replay\_Trace}{trace, spec, journal}
    \State $spec[state] \gets trace[init\_state]$\Comment{initial state of the specification}
    \While{$syscall := shift(trace[syscalls])$}
        \State $real\_result := syscall[result]$
        \State $spec\_result := Replay\_Syscall(spec, syscall)$
        \State \textbf{switch} $(spec\_result,real\_result)$ \textbf{of}
        \State \quad \textbf{case} $(Denied,Granted)$ \textbf{:}
        \State \qquad $journal \gets (CRIT, syscall)$\Comment{An error with its level}
        \State \qquad $\textbf{return} (Failure, journal)$
        \State \quad \textbf{case} $(Granted,Denied)$ \textbf{:}
        \State \qquad $journal \gets Check\_ErrCode(syscall)$
        \State \qquad $spec[state] \gets revert(spec[state],syscall)$\Comment{rollback update}
        \State \textbf{end switch}
    \EndWhile
    \State $journal \gets compare\_states(trace[final\_state], spec[state])$
    \State \textbf{return} $(Success, journal)$
    \EndProcedure
    \end{algorithmic}
\end{algorithm}

We need to check that the result of the replaying conforms the result of the real system trace execution~(see~Alg.~\ref{lst:trace_replay}):

\begin{enumerate}
    \item If access is granted or denied on both the specification and the real system, then we should proceed to the next system call~(lines 3,~14);
    \item If access is granted on the real system, but it is denied on the Event-B specification~(line~7), this clearly signals about an error in the kernel or in the security module. It is not possible to proceed further after this kind of error, the analysis is stopped;
    \item If access is denied on the real system, but it is granted on the Event-B specification~(line~10), the return code of the system call is 
    investigated~(see~Alg.~\ref{lst:check_error_code}). For example, if the return code signals about\footnote{The listed error codes are taken from the Linux kernel file\\ \texttt{include/uapi/asm-generic/errno-base.h}}:
    \begin{itemize}
        \item not enough memory in the system, then no additional actions are taken;
        \item invalid values in the system call's arguments, then with high probability this means that the specification is not complete. This divergence is recorded to the anomaly journal;
        \item not enough permissions. That means there is an error in the kernel or in the security module. This kind of divergence is recorded to the journal.
    \end{itemize}
    After this we restore the previous state of the specification~(line~12) and proceed to the next system call from the trace.
    \item If the replay reaches the end of the trace the global states of the kernel and the specification are compared. The divergences are logged to the anomaly journal.
\end{enumerate}

\begin{algorithm}[h]
    \caption{Investigation of the error code of a system call}\label{lst:check_error_code}
    \begin{algorithmic}[1]
    \Procedure{Check\_ErrCode}{syscall}
    \State $err\_code := syscall[result][err\_code]$
    \State \textbf{switch} $err\_code$ \textbf{of}
    \State \quad \textbf{case} ENOMEM \textbf{:}\Comment{Out of memory}
    \State \qquad \textbf{return} $\varnothing$
    \State \quad \textbf{case} EINVAL \textbf{:}\Comment{Invalid argument}
    \State \qquad \textbf{return} $(WARN, syscall)$
    \State \quad \textbf{case} EACCES \textbf{:}\Comment{Permission denied}
    \State \qquad \textbf{return} $(CRIT, syscall)$
    \State \quad \textbf{case} \dots \textbf{:}
    \State \qquad \dots
    \State \textbf{end switch}
    \EndProcedure
  \end{algorithmic}
\end{algorithm}

The replay analysis outputs the journal of the divergences between the behavior of the real system and the modelled behavior of the Event-B specification.
The journal records need to be analyzed manually to reveal flaws in the specification or the implementation.
However, if no divergences were found then with a certain level of certainty based on obtained sources and specification coverage, we can claim that we successfully demonstrated conformance between the implementation and its specification. 

We measure the coverage by lines of code of the security module and the number of covered LSM hooks across the kernel.
The specification allows more behaviors (states) than it is possible to observe on the real system, thus the specification coverage consists of covered global invariants and different conjuncts of guards conditions.
To evaluate the proposed algorithms we have manually translated a part of the Event-B specification of the HIMACF model to an executable program and tested it on the system call traces gathered with SystemTap.

\section{Related Work}

In~\cite{zanin2004towards} Zanin and Mancini present a formal model for analyzing an arbitrary security policy configuration for SELinux.
At the end of the paper the authors propose an algorithm based on their model for verifying whether, given an arbitrary security policy configuration, a given subject can access a given object in a given mode.
However, they don't go down to the SELinux implementation.

Guttman et al~\cite{guttman2005verifying} present a formalization of the access control mechanism of the SELinux security server together with a labeled transition system representing an SELinux configuration.
Linear temporal logic is used to describe the desired security objectives.
The authors use model checking to determine whether security goals hold in a given system.

There are other examples of using formal methods such as B and TLA+ to formalize and prove correctness of various access control mechanisms or security policy models~\cite{huynh_validating_2014,kozachok_tla+_2018}, but they also do not consider the implementation.

In~\cite{lsm_hooks_runtime_verification} the correctness of LSM hooks placement in the Linux kernel is analyzed. 
The proposed runtime verification method leverages the fact that most of the LSM hooks are correctly placed to find the misplaced ones.

The authors of~\cite{georget2017verifying} analyze the information flows in the LSM framework.
They verify that for any execution path in the kernel starting with a system call and leading to an information flow, there is at least one LSM hook before the flow is performed.
The analysis statically checks the control flow graphs of kernel functions, which are obtained by a compiler plugin during the kernel build, for existence of feasible paths without mediation of the LSM framework.

\section{Conclusion and Future Work}
We have outlined a method for verification of the access control mechanisms implemented as a module inside the Linux kernel for conformance with its abstract specification.
The method consists of several steps.
First, one needs to formalize the specification of the access control mechanisms in the Event-B language and prove its correctness.
Then, since the resulting Event-B specification is high-level and too different from the concrete data structures and functions of the Linux kernel, we propose to develop an additional specification of the Linux system call interface and prove that it conforms to the Event-B specification of access control mechanisms.
Next, we trace system calls to the kernel while performing a series of typical user actions and tests.
Finally, we replay them on the Event-B specification of the system call interface to check the obtained accesses satisfy the security policy model.

We have evaluated the proposed method on the HIMACF model, which integrates several advanced access control mechanisms, and its implementation inside Astra Linux distribution.
We have developed and proved both Event-B specifications, which are required by the method.
We have found that the specification of the system call interface, which is required by the method, turns out to be much larger and more complex than the specification of the security policy model.
A part of the specification was manually translated to an executable form to obtain the proof of concept and test the replay algorithm of the proposed method.
For this we have gathered system call traces with the SystemTap tool.
The future work involves development of a translator from Event-B to an effective executable form and research the possibility of simultaneous OS execution and in-kernel verification of accesses.

\bibliographystyle{splncs04}
\bibliography{refs}

\begin{thebibliography}{10}
\providecommand{\url}[1]{\texttt{#1}}
\providecommand{\urlprefix}{URL }
\providecommand{\doi}[1]{https://doi.org/#1}

\bibitem{abrial_modeling_2010}
Abrial, J.R.: Modeling in {Event}-{B}: {System} and {Software} {Engineering}.
  Cambridge University Press, New York, NY, USA, 1st edn. (2010)

\bibitem{abrial_rodin_2010}
Abrial, J.R., Butler, M., Hallerstede, S., Hoang, T.S., Mehta, F., Voisin, L.:
  Rodin: an open toolset for modelling and reasoning in {Event}-{B}. Int J
  Softw Tools Technol Transfer  \textbf{12}(6),  447--466 (Nov 2010).
  \doi{10.1007/s10009-010-0145-y}

\bibitem{abrial_refinement_2007}
Abrial, J.R., Hallerstede, S.: Refinement, decomposition, and instantiation of
  discrete models: Application to event-b. Fundamenta Informaticae
  \textbf{77},  1--28 (04 2007)

\bibitem{bell_secure_1976}
Bell, D.E., La~Padula, L.J.: {Bell, D. E., LaPadula, L. J. Secure Computer
  System: Unified Exposition and MULTICS Interpretation. ESD-TR-75-306,
  Electronic Systems Division, AFSC, Hanscom AFB, 1976.} (1976)

\bibitem{bell_secure_1973}
Bell, D.E., LaPadula, L.J.: {Secure Computer Systems: Mathematical Foundations.
  ESD-TR-73-278 v. 1, Electronic Systems Division, AFSC, Hanscom AFB} (1973)

\bibitem{viro_lsm_mh_2006}
Belousov, K., Viro, A.: Linux kernel {LSM} file permission hook restriction
  bypass. \url{https://vulners.com/osvdb/OSVDB:25747} (2006)

\bibitem{biba_integrity_1977}
Biba, K.: {Integrity considerations for secure computer systems. Technical
  Report MTR-3153, The MITRE Corporation} (1977)

\bibitem{devyanin_book_2013}
Devyanin, P.N.: The models of security of computer systems: access control and
  information flows. (in Russian). Goryachaya Liniya-Telecom, Moscow, Russia
  (2013)

\bibitem{devyanin_formal_2014}
Devyanin, P., Khoroshilov, A., Kuliamin, V., Petrenko, A., Shchepetkov, I.:
  Formal {Verification} of {OS} {Security} {Model} with {Alloy} and
  {Event}-{B}. In: International Conference on Abstract State Machines, Alloy,
  B, TLA, VDM, and Z. pp. 309--313 (Jun 2014).
  \doi{10.1007/978-3-662-43652-3\_30}

\bibitem{shchepetkov_refinement_2016}
Devyanin, P.N., Khoroshilov, A.V., Kuliamin, V.V., Petrenko, A.K., Shchepetkov,
  I.V.: Using {Refinement} in {Formal} {Development} of {OS} {Security}
  {Model}. In: Mazzara, M., Voronkov, A. (eds.) Perspectives of {System}
  {Informatics}. pp. 107--115. Lecture {Notes} in {Computer} {Science},
  Springer International Publishing (2016)

\bibitem{lsm_hooks_runtime_verification}
Edwards, A., Jaeger, T., Zhang, X.: {Runtime Verification of Authorization Hook
  Placement for the Linux Security Modules Framework}. In: Proceedings of the
  9th ACM Conference on Computer and Communications Security. pp. 225--234. CCS
  '02, ACM, New York, NY, USA (2002). \doi{10.1145/586110.586141},
  \url{http://doi.acm.org/10.1145/586110.586141}

\bibitem{georget_lsm_mh_2016}
Georget, L.: Add missing {LSM} hooks in mq timed {send,receive} and splice.
  \url{http://thread.gmane.org/gmane.linux.kernel.lsm/28737} (2016)

\bibitem{georget2017verifying}
Georget, L., Jaume, M., Tronel, F., Piolle, G., Tong, V.V.T.: Verifying the
  reliability of operating system-level information flow control systems in
  linux. In: 2017 IEEE/ACM 5th International FME Workshop on Formal Methods in
  Software Engineering (FormaliSE). pp. 10--16 (May 2017).
  \doi{10.1109/FormaliSE.2017.1}

\bibitem{goyal_lsm_overlayfs}
Goyal, V.: Overlayfs {SELinux} support. \url{https://lwn.net/Articles/693663/}
  (2016)

\bibitem{guttman2005verifying}
Guttman, J.D., Herzog, A.L., Ramsdell, J.D., Skorupka, C.W.: Verifying
  information flow goals in security-enhanced linux. Journal of Computer
  Security  \textbf{13}(1),  115--134 (2005)

\bibitem{huynh_validating_2014}
Huynh, N., Frappier, M., Mammar, A., Laleau, R., Desharnais, J.: Validating the
  {RBAC} {ANSI} 2012 {Standard} {Using} {B}. In: Abstract {State} {Machines},
  {Alloy}, {B}, {TLA}, {VDM}, and {Z}. pp. 255--270. Lecture {Notes} in
  {Computer} {Science}, Springer, Berlin, Heidelberg (Jun 2014),
  \url{https://link.springer.com/chapter/10.1007/978-3-662-43652-3\_22}

\bibitem{cc1}
{ISO/IEC 15408-1:2009. Information technology -- Security techniques --
  Evaluation criteria for IT security -- Part 1: Introduction and general
  model. ISO} (2009)

\bibitem{cc2}
{ISO/IEC 15408-2:2008. Information technology -- Security techniques --
  Evaluation criteria for IT security -- Part 2: Security functional
  components. ISO} (2008)

\bibitem{jacob2008systemtap}
Jacob, B., Larson, P., Leitao, B., Da~Silva, S.: {SystemTap}: instrumenting the
  linux kernel for analyzing performance and functional problems. IBM Redbook
  \textbf{116} (2008)

\bibitem{jurgens_lsm_infiniband}
Jurgens, D.: {SELinux} support for {Infiniband} {RDMA}.
  \url{https://lwn.net/Articles/684431/} (2016)

\bibitem{kozachok_tla+_2018}
Kozachok, A.: {TLA}+ based access control model specification. Proceedings of
  the Institute for System Programming of the RAS  \textbf{30},  147--162 (Jan
  2018). \doi{10.15514/ISPRAS-2018-30(5)-9}

\bibitem{ltp}
Larson, P.: Testing {Linux} with the {Linux Test Project}. In: Ottawa Linux
  Symposium. p.~265 (2002)

\bibitem{lsm}
Morris, J., Smalley, S., Kroah-Hartman, G.: Linux security modules: General
  security support for the linux kernel. In: USENIX Security Symposium. pp.
  17--31. ACM Berkeley, CA (2002)

\bibitem{astra_se}
{RusBITech}: {Astra Linux{\textregistered} Special Edition}.
  \url{https://astralinux.ru/products/astra-linux-special-edition/}

\bibitem{spruce}
Tsirunyan, K., Martirosyan, V., Tsyvarev, A.: {The Spruce System}: quality
  verification of {Linux} file systems drivers. In: Proceedings of the
  Spring/Summer Young Researchers’ Colloquium on Software Engineering. {ISP
  RAS} (2012)

\bibitem{syzkaller}
Vykov, D.: Syzkaller. \url{https://github.com/google/syzkaller} (2015)

\bibitem{write_lsm_mh_2005}
Write, C.: {LSM} update, another missing hook.
  \url{https://lwn.net/Articles/155496/} (2005)

\bibitem{zanin2004towards}
Zanin, G., Mancini, L.V.: Towards a formal model for security policies
  specification and validation in the {SELinux} system. In: Proceedings of the
  ninth {ACM} symposium on Access control models and technologies. pp.
  136--145. ACM (2004)

\end{thebibliography}

\end{document}